\documentclass{IEEEcsmag}

\usepackage{filecontents}

\usepackage[colorlinks,urlcolor=blue,linkcolor=blue,citecolor=blue]{hyperref}
\expandafter\def\expandafter\UrlBreaks\expandafter{\UrlBreaks\do\/\do\*\do\-\do\~\do\'\do\"\do\-}
\usepackage{upmath,color}

\newcommand\notype[1]{\unskip}
\newcommand*{\nolink}[1]{%
  \begin{NoHyper}#1\end{NoHyper}%
}

\usepackage{listings}
\usepackage{xcolor}

\definecolor{codegreen}{rgb}{0,0.6,0}
\definecolor{codegray}{rgb}{0.5,0.5,0.5}
\definecolor{codepurple}{rgb}{0.58,0,0.82}
\definecolor{backcolour}{rgb}{0.95,0.95,0.92}

\lstdefinestyle{mystyle}{
    backgroundcolor=\color{backcolour},   
    commentstyle=\color{codegreen},
    keywordstyle=\color{magenta},
    numberstyle=\tiny\color{codegray},
    stringstyle=\color{codepurple},
    basicstyle=\ttfamily\footnotesize,
    breakatwhitespace=false,         
    breaklines=true,                 
    captionpos=b,                    
    keepspaces=true,                 
    numbers=left,                    
    numbersep=5pt,                  
    showspaces=false,                
    showstringspaces=false,
    showtabs=false,                  
    tabsize=2
}

\jvol{XX}
\jnum{XX}
\paper{8}
\jmonth{Month}
\jname{IT Professional}
\jtitle{IT Professional}
\pubyear{2024}

\begin{document}

\sptitle{\normalsize This is the accepted version of an article that has been published in this journal. The final version of record is available at https://doi.org/10.1109/MITP.2024.3421968 with OA Under a Creative Commons License}

\title{Enhanced FIWARE-Based Architecture for Cyber-Physical Systems with tinyML and MLOps: A Case Study on Urban Mobility Systems}

\author{Javier Conde}
\affil{Departamento de Ingeniería de Sistemas Telemáticos, Escuela Técnica Superior de Ingenieros de Telecomunicación, Universidad Politécnica de Madrid}
\author{Andr\'es Munoz-Arcentales}
\affil{Departamento de Ingeniería de Sistemas Telemáticos, Escuela Técnica Superior de Ingenieros de Telecomunicación, Universidad Politécnica de Madrid}
\author{Álvaro Alonso}
\affil{Departamento de Ingeniería de Sistemas Telemáticos, Escuela Técnica Superior de Ingenieros de Telecomunicación, Universidad Politécnica de Madrid}
\author{Joaqu\'in Salvachúa}
\affil{Departamento de Ingeniería de Sistemas Telemáticos, Escuela Técnica Superior de Ingenieros de Telecomunicación, Universidad Politécnica de Madrid}
\author{Gabriel Huecas}
\affil{Departamento de Ingeniería de Sistemas Telemáticos, Escuela Técnica Superior de Ingenieros de Telecomunicación, Universidad Politécnica de Madrid}

\markboth{THEME/FEATURE/DEPARTMENT}{THEME/FEATURE/DEPARTMENT}

\begin{abstract}
The rise of artificial intelligence and the Internet of Things is accelerating the digital transformation of society. Mobility computing presents specific barriers due to its real-time requirements, decentralization, and connectivity through wireless networks. New research on Edge Computing and tinyML explores the execution of Artificial Intelligence models on low-performance devices to address these issues. However, there are not many studies proposing agnostic architectures that manage the entire lifecycle of Intelligent Cyber-Physical Systems. This paper extends a previous architecture based on FIWARE software components to implement the Machine Learning Operations flow, enabling the management of the entire tinyML lifecycle in Cyber-Physical Systems. We also provide a use case to showcase how to implement the FIWARE architecture through a complete example of a smart traffic system. We conclude that the FIWARE ecosystem constitutes a real reference option for developing tinyML and Edge Computing in Cyber-Physical Systems.
\end{abstract}

\maketitle

\chapteri{T}he growth of the Internet of Things (IoT) has enabled the deployment of a multitude of devices throughout the world. By 2021 there were 12.2 billion active devices and this number is expected to increase to around 27 billion connected devices by 2025~\cite{state_of_iot_2022}. 
IoT, in combination with other technologies such as Cloud Computing, Big Data, and Aritifical Intelligence (AI) has enabled the expansion of Digital Twins (DTs) and Cyber-Physical Systems (CPS)~\cite{technological_cap}. Both DTs and CPS involve the integration of the digital and physical world through a bidirectional communication: (1) the physical entity collects information from the real world; (2) it transmits this information to the virtual environment; (3) the virtual entity processes the data; (4) the virtual entity sends commands to the physical entity; (5) the physical entity updates its state. Cyber-Physical Systems (CPS) have been applied to all societal activities, including complex systems with specific requirements such as Cyber-Physical Mobile Systems (e.g., smart vehicles) located in dynamic environments (e.g., Smart Cities)~\cite{implementation_of_cps}. 

These systems have a set of requirements that need to be addressed, such as real-time processing, guaranteeing reliability and security, handling large amounts of data, inter-device communication, and data ingestion standardization from heterogeneous environments~\cite{a_survey_of_itnernet}. In the literature, there are proposals aimed at providing solutions to these challenges. Conde et al.~\cite{collaboration_of_digital_twins} propose the FIWARE ecosystem as an open-source solution for the implementation of scalable, AI-driven, and use-case-agnostic DTs and CPS. The FIWARE initiative\footnote{FIWARE: \url{https://www.fiware.org/}} is defined as an ecosystem of software components that facilitate the development of intelligent solutions. Every FIWARE solution is based on communication using the NGSIv2\footnote{NGSIv2: \url{https://swagger.lab.fiware.org/}} (Next Generation Service Interface v2) or NGSI-LD\footnote{NGSI-LD: \url{https://forge.etsi.org/rep/NGSI-LD/NGSI-LD}} (NGSI for linked data) data formats. A central component called Context Broker\footnote{Orion CB: \url{https://github.com/FIWARE/context.Orion-LD}} acts as the center-piece of the solution in charge of managing the context information. It implements the NGSI API and offers synchronous and asynchronous communication capabilities. The other components communicate with the Context Broker providing additional functionalities to the CPS. Table~\ref{table:fiware} summarizes some FIWARE components and their roles in the CPS. The reference architecture for DTs and CPS~\cite{collaboration_of_digital_twins} has been validated in real scenarios such as the Airport of Aberdeen~\cite{applying_Digital}. However, the proposal of Conde et al. processes all the information in the cloud, including the execution of AI models. Consequently, it is necessary to keep IoT devices connected to the rest of the infrastructure, causing high latencies, high server demands, and service loss in case of a communication failure.

\begin{table}
\vspace*{0pt}
\caption{Role of FIWARE Components in the development of CPS}
\label{table:fiware}
\tablefont
\begin{small}
\begin{tabular*}{18pc}{@{}p{78pt}<{\raggedright}p{130pt}<{\raggedright}@{}}
\toprule
Component& Function \\
\colrule
NGSI-LD & Data format homogenization \\ \\
Smart Data Models & Data models homogenization \\ \\
Orion & Real-time communication \\ \\
IoT Agents & Communication with IoT devices\\ \\
Draco & Data ingestion \\ \\
Keyrock and Wilma & Authentication and Authorization \\ \\
Cosmos & Big Data and ML Processing
\end{tabular*}\vspace*{0pt}
\end{small}
\label{tab1}
\end{table}

Other investigations study architectures in which data are processed in a location close to the data sources. This paradigm, known as Edge Computing, has advantages such as reduced latency, energy savings, scalability, and security by not having to send data over the network~\cite{ai_models}. Particular cases of Edge Computing include EdgeAI and TinyML, which involve the deployment of AI models on the edge or directly on low-performance devices~\cite{a_review}.

TinyML investigation has opened new research lines in hardware optimization, software, and 
AI, such as model compression, model quantization, or model pruning. Thanks to tinyML, it is now possible to execute complex models in tiny devices~\cite{TinyML_for_Ultra, 10343114}.

Although tinyML is progressing very fast, it does not mean that IoT devices must be isolated. IoT devices may receive information from the cloud that they are not capable of obtaining or processing. Additionally, the IoT devices may communicate with other systems. There needs to be a balance between processing on the device and processing in the cloud. To achieve this, the tinyML deployment group proposes MLOps (Machine Learning Operations) as an approach to managing the tinyML lifecycle~\cite{tinyML_white_paper}.

The article is structured as follows. In the next section, we extend the architecture based on the FIWARE initiative for DTs and CPS~\cite{collaboration_of_digital_twins} to integrate tinyML and MLOps into it and explain its benefits. In the third section, we apply our proposal to an easy-to-follow use case of a smart traffic barrier. Lastly, conclusions and future work arising from our research are presented. 

\section{EXTENDING THE FIWARE ARCHITECTURE TO INTEGRATE MLOPS LIFECYCLE AND TINYML IN CPS}


Obtaining efficient Machine Learning Models is not the only task in tinyML. Kolltveit and Li~\cite{operationalizing} differentiate between two phases, development, and operationalization. In the development phase, the ML model is trained and evaluated. It requires batch processing of large amounts of data. In the operationalization phase, the model is loaded in the final device that is the one that executes it. The MLOps methodology establishes a fine-grained division of all tasks involving the ML lifecyle. According to Amine et al.~\cite{servless_on_machine_learning} the MLOps process includes (1) data retrieval, (2) data preparation, (3) model training, (4) evaluation, (5) tuning, (6) deployment, and (7) monitoring. These phases coincide with the requirements defined by the tinyML Foundation~\cite{tinyML_white_paper} regarding tinyML lifecycle.

\begin{figure}
\centerline{\includegraphics[width=17.5pc]{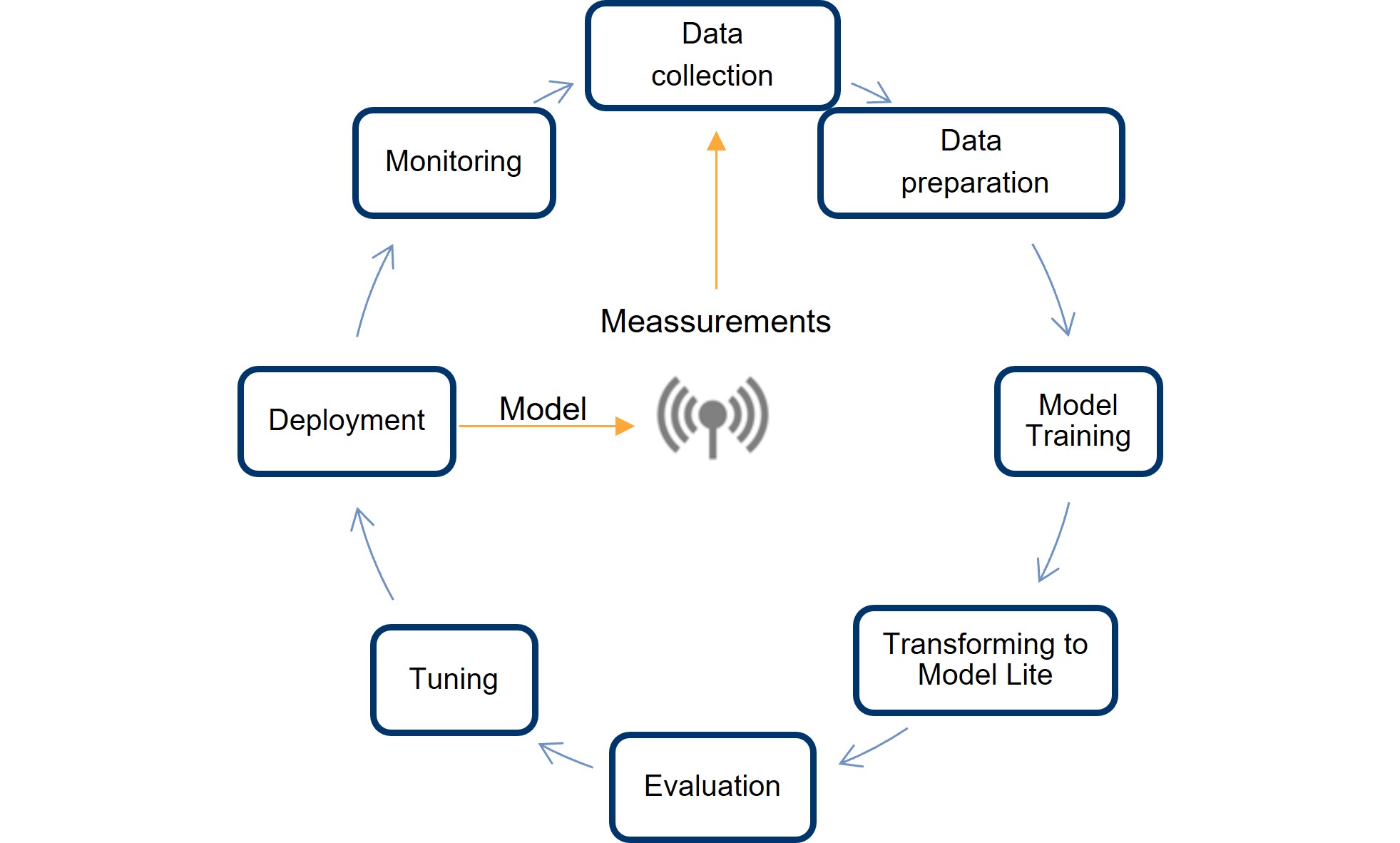}}
\caption{MLOps cycle adapted to tinyML}\vspace*{-5pt}
\label{fig:mlops}
\end{figure}

\nolink{\textbf{Figure \ref{fig:mlops}}} shows the MLOps cycle adapted to the tinyML scenario, including the transformation from the traditional models to their lite version, capable of being run on low-performance devices without excessive loss of the prediction quality. 

In the next subsections, we present the result of our investigation showing how the FIWARE architecture for DTs and CPS can be extended to implement the MLOps cycle and adapt it to the tinyML scenario. We will explain the benefits of our proposal and how it overcomes the barriers mentioned in the literature~\cite{a_survey_of_itnernet}.

\subsection{Collecting data from IoT devices}

The first phase consists of generating historical data to train ML models. The historical data comes from measurements of the devices themselves, so it is necessary to define an architecture for data collection, data homogenization, and data storage. According to Nikolaos et. al~\cite{TinyML_for_Ultra} data collection from sensors is a required task when there are not specific datasets to the use case, or there are not datasets with the level of quality needed.

FIWARE IDAS offers a set of components called IoT Agents that facilitate the integration of external systems, including IoT devices into the FIWARE ecosystem. IoT Agents act as middleware between the Context Broker and the physical device. That is, they translate NGSI-LD HTTP requests coming from the Context Broker into the native data format and protocol of the IoT device. The Context Broker communicates with the IoT device through commands processed by the IoT Agent, and the IoT Agent communicates with the Context Broker by updating the properties of the corresponding NGSI-LD entity. 

IoT Agents ease the dataset generation but also the deployment phase, which is transparent to the IoT devices. The information will be managed by the Orion-LD Context Broker. This proposal also solves the problem of integrating heterogeneous sources in terms of communication protocols, simplifying the rest of the MLOps phases as all the data will be accessible through the NGSI-LD format.

\subsection{Modeling and storing data}

Data in the CPS are modeled through NGSI-LD, which is agnostic to the use case. The FIWARE Smart Data Models initiative\footnote{Smart Data Models: \url{https://smartdatamodels.org/}} defines specific data models for diverse smart domains. In this sense, NGSI-LD allows homogenizing data formats, while the FIWARE Smart Data Models allow homogenizing data models. Both are essential for complex systems such as CPS, where many actors are involved.

FIWARE Draco \footnote{FIWARE Draco: \url{https://github.com/ging/fiware-draco}}, based on Apache Nifi, is a high-scalable tool for transforming and routing data in real-time. It is used to integrate systems different from IoT devices in the CPS and to generate historical data for the training phase.  

\subsection{Training Data and Managing ML Versions}

 The training phase can begin once the historical data are available. Different frameworks and libraries adapted to tinyML scenarios are mentioned in the literature (e.g., TensorFlow Lite, uTensor, Pythorch Mobile). One of the most widely used is TensorFlow Lite\footnote{TensorFlow: \url{https://www.tensorflow.org}} because of its efficiency in compressing models and adapting them to a wide variety of microcontrollers. TensorFlow Lite is focused on Neural Networks. Emlearn\footnote{Emlearn library: \url{https://github.com/emlearn/emlearn}} is a library that converts the model trained through the Keras\footnote{Keras Library:\url{https://keras.io/}} or Scikit-learn\footnote{Scikit-learn library: \url{http://scikit-learn.org/}} libraries to C code. Emlearn is compatible with any device that includes a C99 compiler. It has been tested on microcontrollers such as ESP32, ESP8266, or STM32.

Machine learning does not follow a linear cycle, it is an iterative process. Better models can be obtained by tuning the hyperparameters, changing the algorithm, increasing the quality and/or quantity of historical data, etc. We have identified the necessity of ML orchestrators that help to monitorize and automatize all phases of the MLOPs life cycle. Open source tools such as MLFlow\footnote{MLFlow: \url{https://mlflow.org/}} or Neptune\footnote{Neptune: \url{https://mlops.neptune.ai/}} allow to manage the different phases of MLOps. These types of tools ease the deployment, tracking, and monitoring of Machine Learning models in IoT devices, as well as support reuse, collaboration, maintenance, and replicability of the scenario. 

CPS are complex systems with many components participating in it. Consequently, we have also identified the necessity of high-level orchestrators that control the entire scenario. In contrast to ML orchestrators that only manage the MLOps cycle, high-level orchestrators can manage every process of the CPS. Tools such as Apache Airflow\footnote{Apache Airflow: \url{https://airflow.apache.org/}}, Perfect\footnote{Perfect: \url{https://www.prefect.io/}}, or Flyte\footnote{Flyte: \url{https://flyte.org/}} are candidates for high-level orchestrators.   

As part of our research, we have integrated Apache Airflow and MLFlow in the original FIWARE architecture\footnote{MLOps supermarket use case: \url{https://github.com/ging/fiware-mlops-supermarket}}.

Our proposal enhances the state of the art, where FIWARE-based architectures for CPS with AI capabilities delegate the prediction tasks to the core of the network, resulting in additional latency, network overhead, and the need to connect IoT devices to the cloud~\cite{10152259, collaboration_of_digital_twins}. Furthermore, these proposals do not integrate the ML lifecycle within the architecture, making them less reproducible and less extensible to new model versions. Similarly, other works do explore the integration of tinyML and MLOps~\cite{TinyML_for_Ultra, janapa2023edge}, but they do not include the phase of information retrieval and IoT management, which is essential to complete the MLOps cycle in CPS. Our proposal thus addresses all these barriers in a single architecture that enables the improvement of CPS operation in an automatable, secure, and real-time manner.

\subsection{Deployment and monitoring}

Once an efficient and device-compatible model has been obtained, it is necessary to define a mechanism to deploy it. The IoT Agents can send commands to be executed on the IoT devices and notify them of the existence of a new model. This approach makes it easy to keep the sensor up-to-date and facilitates the automation of the process.

Regarding monitoring, the device can also send its predictions to the Context Broker through its respective IoT Agent. With these results, the quality of the trained model can be evaluated, and the MLOps lifecycle can be fed back by generating new versions of the model.

\section{USE CASE OF A TINYML-MLOps PIPELINE FOR A SMART TRAFFIC SYSTEM}

Given that FIWARE technology is agnostic to the use case and that the Smart Data Models initiative covers a wide range of scenarios~\cite{10152259}, this makes our proposal implementable in any field, including smart agriculture, smart aerospace, smart energy, etc.

As an example of implementation, we present a use case for an automatic traffic barrier system that serves as a reference guide for the integration of the MLOps process for TinyML in CPS mobility systems. This scenario contains two sensors that are not connected to each other. The first detects the density of vehicles in a specific area. The second is an smart traffic barrier that controls traffic of such area.  

The main goal is to train a Machine Learning model and deploy it in the traffic control system. The model predicts the density of vehicles in the area, and based on the result obtained, the smart traffic barrier allows vehicles to pass or not. The scenario is dockerized so it can be easily replicated and deployed. All the code is available in a public repository\footnote{FIWARE TinyML and MLOps traffic use case: \url{https://github.com/ging/tinyML-MLOps-Fiware-Barrier}}.

\begin{figure*}
\centerline{\includegraphics[width=25pc]{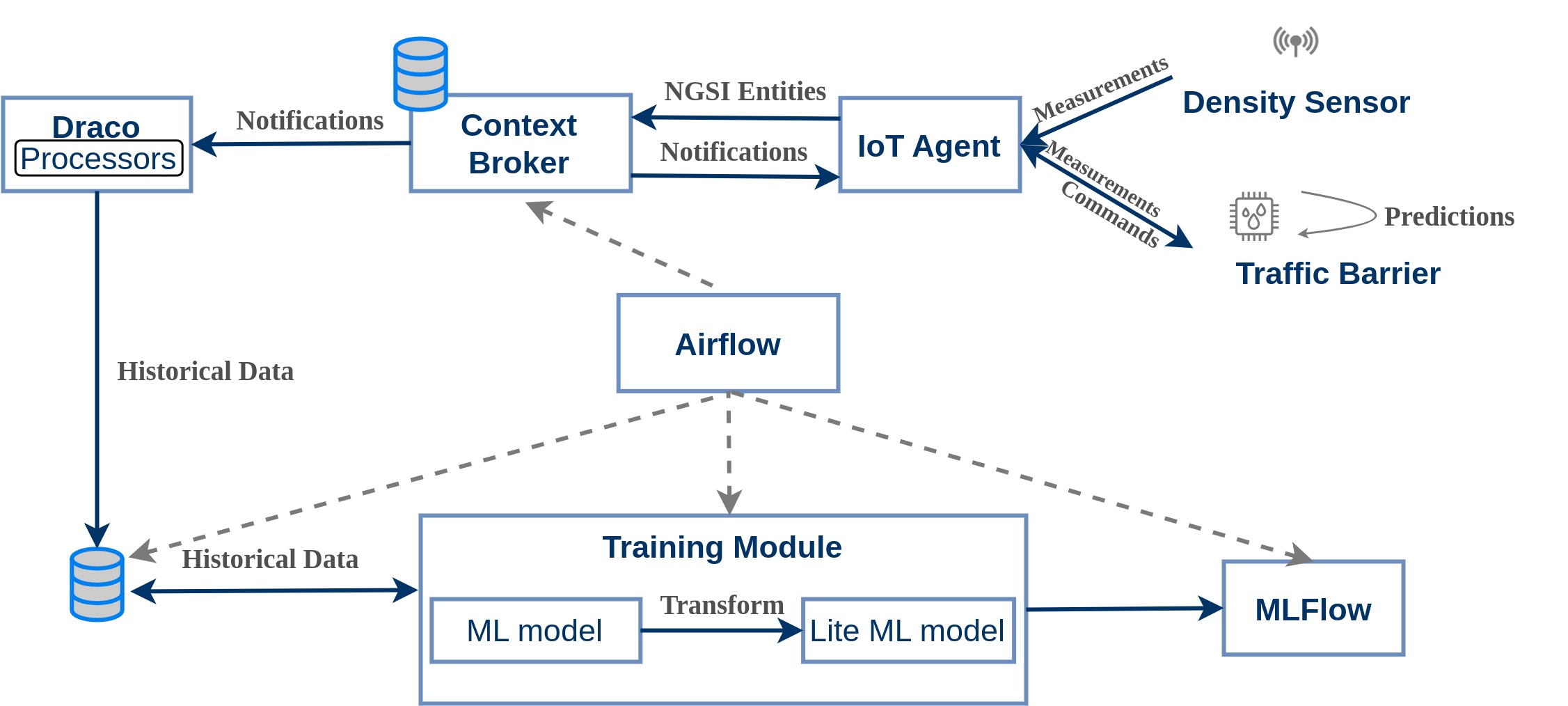}}
\caption{General Architecture of the smart traffic system}\vspace*{-5pt}
\label{fig:arch}
\end{figure*}

\nolink{\textbf{Figure \ref{fig:arch}}} shows the FIWARE component-based architecture for TinyML-MLOps lifecycle adapted to the traffic system use case.

\subsection{Data Collection and Generation of the Dataset}
In the pre-training phase, it is necessary to generate historical data if it they do not exist previously. In this scenario data were collected from a loop situated on a road of Santander city (Spain) which records the density of vehicles (vehicles/minute) passing over it. The dataset also includes meteorological information from Santander provided by the Spanish government. Data were collected throughout half of the year of 2021, with one measurement every 15 minutes. Draco was used to generate the historical data. Upon changes in real-time NGSI-LD entities, Draco receives notifications, processes them, and stores the results in a Mongo database, which was used to generate the training CSV.





\subsection{Training the density of vehicle model}

Once the information is stored in Mongo, the data are cleaned and processed. The predictor uses the environment data, date, and time information to train a model that classifies the density as low, when there are less than ten vehicles per minute; and high, in the opposite case. The binary classifier is a random forest implemented by the scikit-learn library. Firstly, we trained a model named large\_model, configuring the RF with 50 trees and a maximum depth of 10, resulting in a model accuracy of 91.4\%. This model took 301.83 ms to train and 9.18 ms to make 3,267 predictions (with an average of 2.81 $\mu$s per prediction), and it occupies 2.79 MB. The resultant model is transformed by the emlearn library so that it can be deployed and executed on the barrier controller. The tiny model was compressed by reducing the RF to 10 trees, with a maximum depth of 8 levels, achieving an accuracy of 89.7\% (1.7\% lower than the uncompressed model). It took 65.54 ms to train and transform the model (78.3\% less time), 1.61 ms to make 3,267 predictions (with an average of 0.49 $\mu$s per prediction), and occupies 0.13 MB (95.35\% smaller than the large\_model.), making the model fit within an ESP32 with 4MB of memory. The training process is managed within MLFlow. MLFlow generates metadata of the process such as the algorithm, the hyperparameters, the link to the historical data, the evaluation results, or the version. The outcome of this module is a new version of the Machine Learning model ready to be deployed on the smart traffic barrier.

\subsection{Execution of the model in the smart traffic barrier}  

Once the data model has been generated it can be uploaded to the smart traffic barrier and it can start operating. In a traditional approach, the prediction would be made in the cloud and the result would be transmitted to the smart traffic barrier. In contrast, with our proposal based on tinyML, the model is loaded onto the IoT device, which executes the prediction internally, saving battery, time, and without interacting with other systems.

The whole process is automated through Airflow. A Direct Acyclic Graph (DAG) schedules a new ML cycle by training and deploying a new tinyML model when new data are collected. In our scenario, the model was retrained with a dataset covering all of 2021 instead of just half the year. The newly compressed model improved its accuracy to 93.1\%, took 103 ms to train, and occupies 0.13 MB.

\section{CONCLUSION AND FUTURE WORK}
The evolution of IoT has motivated the trend of bringing computing closer to the edge of the network, including the execution of Machine Learning models. Most tinyML research focuses on hardware improvement and the generation of efficient models to transfer processing to the IoT device. However, a complete architecture that performs all prerequisites through MLOps must be implemented. In this article, we propose to extend the FIWARE ecosystem, which has already been validated to implement DTs and CPS, by incorporating tinyML and MLOps capabilities. We explained the benefits of our proposal, which fulfills the requirements for CPS deployed in heterogeneous environments. Moreover, this article has a practical approach, including a use case of a smart traffic system. From our work, we can conclude that we have enhanced the FIWARE capabilities for developing CPS by integrating tinyML capabilities managed by the MLOps life cycle.

For future research, the architecture can be extended with new MLOps and high-level orchestrators, and it will be validated within new use cases from different sectors. Furthermore, the architecture presented can be distributed through a federated Context Broker with the aim of bringing the training phase as close as possible to the IoT device. Another future line is related to the integration of Large Language Models (LLMs) on edge devices. New research should focus on optimizing LLMs, ensuring an automatic, fast and accurate deployment in CPS while minimizing resource demands.

\vspace*{-8pt}

\section{ACKNOWLEDGMENTS}
This work was supported by the FUN4DATE (PID2022-136684OB-C22) project funded by the Spanish Agencia Estatal de Investigación (AEI) and by the Chips Act Joint Undertaking project SMARTY (Grant no. 101140087).

\def\refname{REFERENCES}

\bibliographystyle{IEEEtran}
\bibliography{IEEEabrv, bibliography}

\begin {IEEEbiography} {Javier Conde}{\,} is an Assitant Professor with the UPM. His research interests lie in the fields of Digital Twins and Machine Learning.
\end{IEEEbiography}

\begin{IEEEbiography}{Andr\'es Munoz-Arcentales} is currently an Assitant Professor with the UPM. with a major research interests in Data Engineering, and Data Usage Control.
\end{IEEEbiography}

\begin{IEEEbiography}{\'Alvaro Alonso}{\,} is currently an Associate Professor with the UPM. His research interests include Multi-videoconferencing Systems, and Security Management.
\end{IEEEbiography}

\begin{IEEEbiography}{Joaqu\'in Salvach\'ua}{\,} is currently an Associate Professor with UPM. His research interests include Cloud and Edge architectures.
\end{IEEEbiography}

\begin {IEEEbiography}{Gabriel Huecas}{\,} is currently an Associate Professor with UPM. His research interests include Digitization, and Data Spaces.
\end{IEEEbiography}

\end{document}